# Decomposition, diffusion, and growth rate anisotropies in self-limited profiles during metalorganic vapor-phase epitaxy of seeded nanostructures

E. Pelucchi
*Tyndall National Institute, "Lee Maltings," Dyke Parade, Cork, Ireland*
*Laboratory of Physics of Nanostructures, Ecole Polytechnique Fédérale de Lausanne, CH-1015 Lausanne, Switzerland*

V. Dimastrodonato
*Tyndall National Institute, "Lee Maltings," Dyke Parade, Cork, Ireland*

A. Rudra, K. Leifer[*], and E. Kapon
*Laboratory of Physics of Nanostructures, Ecole Polytechnique Fédérale de Lausanne, CH-1015 Lausanne, Switzerland*

L. Bethke, P. Zestanakis, and D. D. Vvedensky
*The Blackett Laboratory, Imperial College London, London SW7 2AZ, United Kingdom*
(Dated: March 11, 2011)

We present a model for the interplay between the fundamental phenomena responsible for the formation of nanostructures by metalorganic vapour phase epitaxy on patterned (001)/(111)B GaAs substrates. Experiments have demonstrated that V-groove quantum wires and pyramidal quantum dots form as a consequence of a self-limiting profile that develops, respectively, at the bottom of V-grooves and inverted pyramids. Our model is based on a system of reaction-diffusion equations, one for each crystallographic facet that defines the pattern, and include the group III precursors, their decomposition and diffusion kinetics (for which we discuss the experimental evidence), and the subsequent diffusion and incorporation kinetics of the group-III atoms released by the precursors. This approach can be applied to any facet configuration, including pyramidal quantum dots, but we focus on the particular case of V-groove templates and offer an explanation for the self-limited profile and the Ga segregation observed in the V-groove. The explicit inclusion of the precursor decomposition kinetics and the diffusion of the atomic species revises and generalizes the earlier work of Basiol et al. [Phys. Rev. Lett. **81**, 2962 (1998); Phys. Rev. B **65**, 205306 (2002)] and is shown to be essential for obtaining a complete description of self-limiting growth. The solution of the system of equations yields spatially resolved adatom concentrations, from which average facet growth rates are calculated. This provides the basis for determining the conditions that yield self-limiting growth. The foregoing scenario, previously used to account for the growth modes of vicinal GaAs(001) during MOVPE and the step-edge profiles on the ridges of vicinal surfaces patterned with V-grooves, can be used to describe the morphological evolution of any template composed of distinct facets.

PACS numbers: 68.55.-a, 68.65.-k, 81.05.Ea, 81.10.Aj, 81.10.Bk

## I. INTRODUCTION

There has been considerable effort in recent years aimed at controlling the growth of self-organized semiconductor nanostructures, driven by the broad range of potential applications, from conventional optoelectronic devices to quantum information processing. A substantial part of this endeavor has been devoted to improving our understanding of the self-assembly of quantum dots (QDs) during the Stranski-Krastanow (SK) growth mode in an attempt to attain site control and thereby reduce the width of their lateral size distributions.[1,2] Control over the growth process becomes essential, for example, when precise spatial and spectral overlap between a single quantum dot or wire and a quantum electrodynamical nanocavity is necessary,[3,4] or when the implementation of reproducible QDs coupling is required.[5]

Seeded metalorganic vapor-phase epitaxy (MOVPE) of nanostructures on pre-patterned semiconductor substrates has emerged in the last 20 years as an important tool for fabricating site-controlled nanostructures[6]. One-dimensional V-groove quantum wires (QWRs) have found successful applications in fundamental studies of one-dimensional transport and as components of optoelectronic devices.[7–9] Inverted pyramidal QDs[6,10] have been shown to provide a valuable alternative to self-assembly when high quality ordered arrays of semiconductor QDs are required. They naturally provide site control, have shown the highest uniformities to date (1–8 meV optical full width at half maximum,[11–16] without detriment effects to the excitonic feature quality), have been demonstrated as efficient single and entangled photon sources in the near infrared spectrum,[17,18] and can be incorporated into single-photon/single-dot *electrically pumped* light emitting devices.[19]

---

[*]Present address: Electron Microscopy and Nanoengineering, Department of Engineering Sciences, Uppsala University, Sweden.

Despite the broad interest in this class of semiconductor nanostructures, several aspects of their fundamental growth principles remain to be fully elucidated. Although the role of capillarity in the formation of V-groove QWRs and QDs has been widely explored and demonstrated,[20,21] comparatively little attention has been given to another fundamental observation: the origin of growth rate anisotropies (GRAs) between the various crystallographic planes during MOVPE on patterned substrates. The importance of GRAs is due to the fact that they are responsible for the AlGaAs/GaAs self-limited profiles, which act as nano-templates for QWR and QD growth. The absence of such a well-defined self-limited profile – as, for example, is the case for molecular-beam epitaxy (MBE) on (100) V-grooved templates – prevents proper lateral confinement and therefore preempts the possibility of obtaining QWRs and QDs.[21,23] GRAs are especially pronounced for pyramidal QDs on (111)B substrates, so much so that very little growth has been observed on (111)B surfaces. The term "no growth" (111)B surfaces was introduced in the early literature to describe this effect, but with no explanation of its microscopic origin.[24,25].

Early attempts at identifying the origin of GRAs in V-groove QWRs were based predominantly on "MBE-like" models with adatom attachment anisotropies on the different crystallographic facets (often linked to Wulff-like constructions) as a possible source of self-limited profiles.[20,26,27] Lelarge et al.[28] observed that large diffusion lengths of the adsorbed species (of the order of tens of microns) must be postulated to justify GRAs in V-grooves,[29] in striking contrast to the reported 200 nm range of adatom diffusion lengths in such systems.[20,28] Decomposition anisotropies were also proposed as a possible source of the GRAs (and of the observed dependence of GRAs on growth temperature), without providing either further theoretical arguments or experimental evidence.[21,28]

Kaluza et al.[30] observed that the self-limited profiles in V-grooves QWRs (and, consequently, the GRAs) are strongly dependent on the organometallic precursor for Ga and Al used in MOVPE. They suggested that the differences could originate from the different decomposition rates of the different precursors on the vicinal (111)A GaAs surfaces. Moreover, some of the present authors[31] have recently found that the anisotropy in the decomposition of metalorganic precursors on (111)A and (111)B surfaces plays a major role in a peculiar and unexpected GRA phenomenon which was observed during the formation of (In)GaAs/AlGaAs pyramidal QDs. By using a substrate pattern consisting of one pyramid placed at the center of a triangular (or hexagonal) area free of pyramids, located within a uniform array of pyramidal recesses, the emission wavelength of the isolated QDs was reproducibly blue shifted with respect to the QD array emission wavelength, i.e. the isolated QD was consistently thinner than the array of QDs.[32,33]

Apart from these recent observations, there is still no general consensus about the role of precursors and/or adatoms as the source of GRAs during seeded MOVPE of nanostructures on pre-patterned semiconductor substrates.[34] We provide in this article additional experimental evidence and theoretical insights that demonstrate that the GRAs (including the "no growth" (111)B surfaces) and the self-limited profile are, in fact, inherently linked to the anisotropies in the surface decomposition processes of the precursors on the different crystallographic facets, both for V-groove QWRs and pyramidal QDs. Moreover, we propose that it is precisely the interplay between these anisotropies and the capillarity fluxes that determines the observed variation of (111)A vicinal facets with growth temperature in V-grooves.

The outline of this paper is as follows. In Sec. II we summarize GRAs in pyramidal quantum dots. Although our main interest here is QWRs in V-grooves, the pronounced anisotropy for QDs on patterned (111)B substrates merits a separate discussion. A preliminary, largely phenomenological, analysis of the temperature-dependent GRAs of QWRs in V-grooves is presented in Sec. III. The main results of this paper are contained in Sec. IV, where we develop our atomistic theory of self-limiting growth based on reaction-diffusion equations for each facet. This theory demonstrates how GRAs arise from facet-dependent decomposition, diffusion, and incorporation kinetics, and can be applied to any surface composed of distinct facets. The extension of these ideas across an extended temperature range is discussed in Sec. V, where we carry out an analysis in the spirit of that in Sec. II, but by using a coarse-grained set of reaction-diffusion equations. This enables us to account qualitatively for the temperature variation of the V-groove sidewall angles formed by the bounding facets while bypassing a full parametrization of our model. Our conclusions and future directions are summarized in Sec. VI.

## II. PRECURSOR DECOMPOSITION: PATTERNED *VERSUS* PLANAR (111)B SUBSTRATES

As noted in the introduction, GRAs are especially pronounced in pyramidal QDs on (111)B substrates. In this section, we present a preliminary discussion of GRAs and their possible causes through experimental observations in pyramidal QDs. The following sections will focus exclusively on V-groove QWRs. Figure 1(a) shows a scanning electron microscope (SEM) plan view of a partially patterned pyramidal QD sample[11,12] after standard growth of less than 0.5 $\mu$m (nominal) of $Al_xGa_{1-x}As$, with $x = 0.30, 0.55, 0.75$, by low pressure MOVPE at 0.5 $\mu$m/hr at an estimated surface temperature $T_g \sim 710°C$. Three distinct areas on the substrate can be identified. On the right of the figure is a regularly patterned area (I) with inverted pyramidal recesses (5 $\mu$m pitch), only partially filled during growth. On the left, a planar unprocessed area (II) is characterized by irregular three-dimensional (3D) growth (enlarged in the inset of Fig. 1(a) is an SEM cross section which shows the defect-rich 3D growth mode). In between these regions is a planar unprocessed area (III) showing a mirror-like surface over approximately 100 $\mu$m, with only a few defects appearing. Cross-sectional SEM and atomic force microscopy (AFM) (not shown) confirmed that no significant epitaxial layers grew in this area.[35] Figures 1(b,c), on the other hand, show that the boundary pyramids (c) are filled to a greater extent (i.e. thicker $Al_xGa_{1-x}As$ layers grew in them) than the pyramids located well inside the patterned regions (b). This transition in pyramid filling (from nearly closed to open) occurs continuously over several tens of microns (not shown).

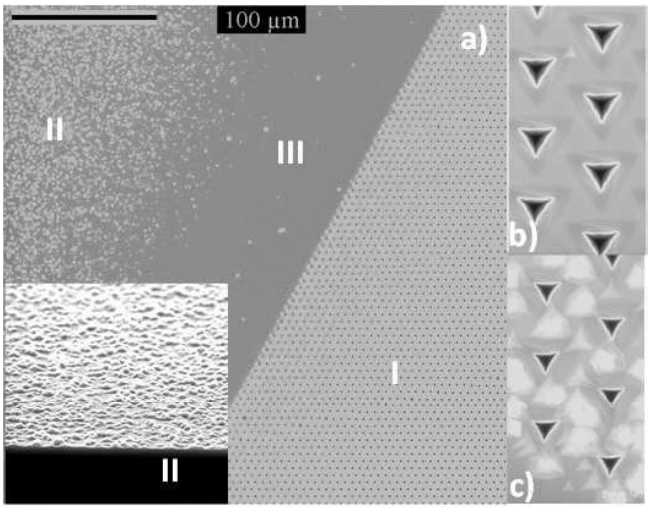

FIG. 1: (Color online) SEM plan view after the growth of QDs on a (111)B substrate patterned with inverted pyramids. (a) Three regions on the substrate: I, a regularly patterned area with partially filled inverted pyramidal recesses (5 $\mu$m pitch), II, a planar unprocessed area with irregular 3D growth, and III, a planar unprocessed area with a mirror-like (no-growth) surface. Inset in (a): Enlarged cross section of region II; (b) SEM image of pyramids located inside the patterned region, and (c) SEM image of pyramids located near the boundary of the patterned region.

Anisotropies in the decomposition process of metalorganic precursors give a simple, appealing and straightforward explanation of these observations. The decomposition rate of the precursors (TMGa/Al) is extremely low on (111)B surfaces and allows for long diffusion lengths prior to decomposition (more than several tens of microns).[13,14,32,36] Under these growth conditions, the decomposition probability of TMGa/Al is so low on the planar (111)B surfaces that nucleation would occur preferentially at defect sites (from which the anomalous 3D growth occurs in region II). On the other hand, the decomposition probability is significantly higher on (111)A surfaces, transforming the patterned areas into efficient traps for the precursors, so that the gradient in the TMGa/Al concentration would induce a net flux of precursors from the planar (111)B surfaces towards the patterned areas. The "no growth" area III is the (111)B surface, which has been depleted by the net flux of precursors towards the patterned area.

This increased flux will also increase the effective growth rate in the "boundary" pyramids, while having little effect on the growth rate deep inside the patterned area. The preferential precursor decomposition inside the pyramidal structure would then deplete the planar (111)B surfaces around the pyramids ("no growth" on the (111)B surface) and provide the excess adatom density necessary for the enhanced growth rate of the (111)A planes inside the pyramids. Following the arguments of Ref. [21] for V-groove structures, (a sketch exemplifying this argument is presented in Fig. 9), whereby a higher "vertical" growth rate on the vicinal (111)A planes relative to the growth rate of the bottom (100)/(111)B facet stabilizes a self-limited profile, we can easily account for the formation of the observed self-limited profile inside the pyramidal structures.

Such significant diffusion lengths observed in Fig. 1 cannot be reconciled with the small adatom diffusion lengths (<1 $\mu$m) reported for Al and Ga atoms on (111) surfaces.[21] The poor decomposition rate of TMGa/Al on (111)B surface also easily resolves the apparent contradiction reported in the literature between the high quality epitaxial layers obtained by MBE on (111)B surfaces with "standard" growth conditions,[37] and the extremely poor (3D and defected) $Al_xGa_{1-x}As$ quality obtained by MOVPE on the same substrates, which is only partially alleviated for growth at extremely high temperatures (800–900°C), i.e. in conditions where the precursor decomposition rates should be substantially enhanced.[13,38,39]

While it is not the purpose of this paper to examine the microscopic origin of this striking decomposition differences, a few general comments can be made. Refs [40] and [41] argue that crystal sites with low coordination numbers (e.g. an edge of an island or step) could be preferential sites for precursor decomposition. Our observations would imply fewer decomposition sites on the (111)B surface than on the (111)A, at least under the growth conditions that we used. In general, the issue of precursor stability and decomposition rate should be addressed with ab initio methods that are capable of comparing the bond stability of the precursors and of different III-V surface configurations (and surface dangling bond arrangements). However, while such calculations can be carried out for simple molecules[42], the decomposition kinetics of the complex group-III precursors used in MOVPE remains a challenge.

Moreover, we observe that gas-phase diffusion models,[43] frequently proposed when long range effects are observed in MOVPE growth on patterned substrates, cannot explain the observations in Fig. 1. Such a model would predict a decreased growth in the pyramidal structures and an increase in the growth rate on the planar (111)B surfaces close to the boundary with the patterned

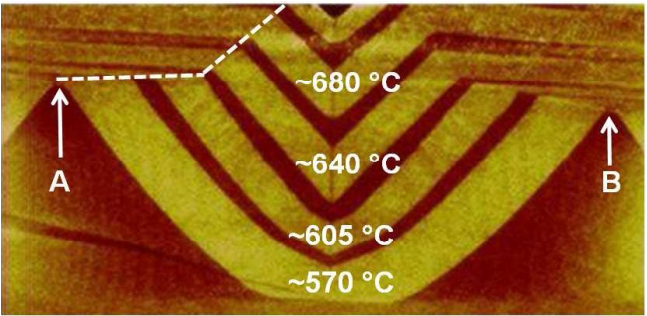

FIG. 2: (Color online) Representative AFM cross-sectional image of an AlGaAs/GaAs multilayer grown by MOVPE at the indicated temperatures on a substrate with a 3 $\mu$m-pitch substrate from A to B. The evolution of the boundary between the (100) and (111)A planes is indicated by the dotted line. The contrast between AlGaAs and GaAs is given by the different height signal of the measuring AFM tip, due to a thin oxide layer grown upon AlGaAs.

area, and is simply inappropriate to describe the fully depleted region III[44].

## III. PRECURSOR DECOMPOSITION AND GROWTH RATE ANISOTROPIES IN V-GROOVE QUANTUM WIRES

Striking effects of GRAs are also observed during AlGaAs/GaAs growth on V-grooved templates.[21] Figure 2 shows a representative AFM cross-sectional image of the epitaxial layers after MOVPE growth on a 3 $\mu$m-pitch substrate. After an initial 400 nm $Al_{0.3}Ga_{0.7}As$ buffer (the thickness values are nominal, as calibrated on a planar (100) substrate), the epitaxial layers were repeated in pairs of 200 nm $Al_{0.3}Ga_{0.7}As$ and 100 nm GaAs, and only the growth temperature was varied from one period to the other. The growth rate was $\sim 1$ $\mu$m/hr, the V/III ratio $r_{V/III}$ $\sim$230 for $Al_{0.3}Ga_{0.7}As$ and $r_{V/III}$ $\sim$180 for GaAs. As indicated in the figure, the estimated surface temperature was ramped from $\sim$570°C to $\sim$680°C.

Several differences between the nominally similar periods can be observed. Here, we concentrate on the differences between the growth rates on the (100) ridge and on the (111)A vicinal planes. To highlight the effect, we have marked the evolution of boundaries between the two layers on the left side of the figure. Clearly, at low temperature, the growth rate on the (100) planes is minimal, with the entire growth process concentrated inside the V-groove. At the lowest temperatures this results in a significant "lateral expansion" of the (100) facet, which tends to quickly "close" ("planarize") the patterned area. In this regime, the GRAs are evident. On the other hand, a sudden change in the planarization appears near $\sim$640°C. The profile between the (100) ridge and the (111)A vicinal planes becomes steeper and the GRAs reduce significantly, but nevertheless maintain a higher growth rate on the (111)A vicinal planes.

The picture of an "anisotropy" in the decomposition rate of precursors advanced in Sec. II again provides insight into the growth process.[45] The decomposition rate $\Gamma_k$ of physisorbed precursors on a given GaAs plane ($k = (100)$ or $k = (111)A$) is expected to follow Arrhenius behavior: $\Gamma_k \sim \exp(-E_k/k_B T_g)$, in which $E_k$ is the energy barrier to decomposition and $k_B$ is Boltzmann's constant.[46] A higher decomposition barrier for the group-III precursors on (100) planes would immediately explain the experimental observations. In fact, at temperatures for which the distribution of precursor energies is mostly below the barrier for decomposition on (100) planes, precursors would simply diffuse towards the (111)A vicinal planes, whereupon they would decompose, releasing the group-III adatoms. These would migrate on the (111)A vicinal planes only for a relative short distance[21] prior to incorporation. Conversely, at higher temperature more precursor molecules are likely to be energetically above both the Arrhenius decomposition barriers, resulting in higher decomposition and growth rates on (100) ridge planes and in a diminished growth rate anisotropy.

Inspection of Fig. 2 shows that the growth on the (100) planes at the bottom of the V-groove (the flat bottom is a consequence of the patterning procedure) at low temperatures is actually significant. This also tends to exclude adatom kinetic anisotropies as being relevant to the process. It would be extremely difficult to understand why an adatom would be more likely to attach to the (100) planes if they reside at the bottom of a V-groove, while totally disregarding the (100) planes on the ridge of the V-groove if a substantial adatom-induced GRA were the source of the observations. The decomposition anisotropy picture, on the other hand, can easily account for the effect, as normal adatom diffusion and capillarity fluxes[20,21,28] would accentuate the migration of adatoms to the center of the V-groove from the region of high concentration of adatoms, i.e. the (111)A planes. If an insignificant growth anisotropy (related to the adatoms) would be present, nothing would then hinder adatom attachment to the bottom (100) planes.

The differences in the decomposition barriers on the (100) and (111)A planes can be estimated by invoking the Arrhenius picture discussed above. If the precursors are assumed to have sufficiently long diffusion lengths[28] and the growth rate on each plane is proportional to the rate of material decomposed *in situ*,[47] then

$$\frac{R_{(100)}}{R_{(111)}} \propto \frac{\Gamma_{(100)}}{\Gamma_{(111)}} \propto \exp\left[\frac{E_{(111)} - E_{(100)}}{k_B T}\right], \quad (1)$$

with $R_{(100)}$ [resp., $R_{(111)}$] being the growth rate perpendicular to the (100) plane [resp., vicinal (111)A planes] of the ridge [resp., the sidewalls]. The assumption that the growth rate depends only on the decomposition rate, while disregarding the effects of diffusion, is based on experimental observations. AFM and transmission electron microscopy (TEM) cross-sectional images (not reported here) show that each crystallographic facet is flat over lengths of $\mu$m. Adatom diffusion covers much smaller lengths (of order of hundreds of nm), so neglecting diffusion in this analysis, where we only consider the constant growth rate in the middle of each facet, can be regarded a valid approximation.

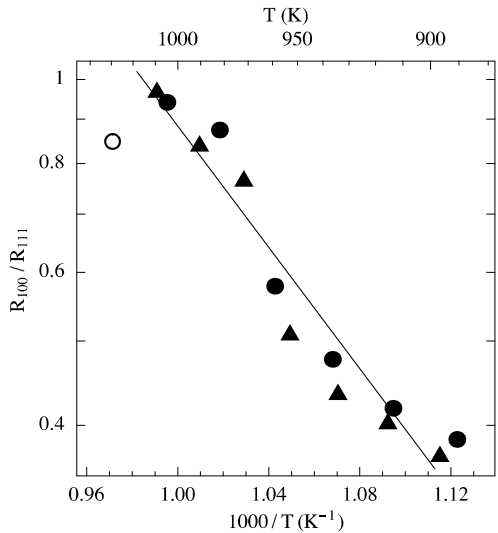

FIG. 3: Arrhenius analysis of the ratio between (100) and (111)A corresponding layers thicknesses of the $Al_{0.3}Ga_{0.7}As$ as from description in the text. The different symbols refer to the two different samples shown in Fig. 4 (triangles refer to sample a) and circles to sample b)). The data point indicated by the open circle was excluded from the fit.

In Fig. 3 we present a comparison of ridge *versus* sidewall thicknesses for two growth runs where (after a thin GaAs buffer) $Al_{0.30}Ga_{0.70}As$ layers (200 nm nominal, $r_{V/III}$ ~220) separated by GaAs markers (10 nm nominal, $r_{V/III}$ ~120) were grown at different temperatures in a (3 $\mu$m-pitch array) V-grooved template. The nominal layer thicknesses (200 nm) and growth interruption have been carefully chosen (based on our experience) to minimize transient effects, which act only in the first $\approx$ 10 nm in this particular sample design. Figure 4 shows two representative SEM images used for the analysis. The sidewalls thicknesses were measured perpendicular to the vicinal crystallographic plane direction, neglecting the minimal angle differences between the planes. $T_g$ was varied in the range $\sim$620°C to $\sim$760°C. The slope of the fit is consistent with a decomposition barrier on the (100) planes that is $0.7 \pm 0.06$ eV higher than on the (111)A planes. A similar analysis for the sample in Fig. 2, where only a few experimental points can be usefully analyzed, so a correspondingly diminished reliability should be expected, yields $0.88 \pm 0.08$ eV.

We stress that this analysis (and our estimate of the difference in decomposition barrier between the crystallographic planes) should be considered only as an approximation. Our simplified model does not completely take into consideration the effective competition between the crystallographic planes during the decomposition/diffusion process. This should be considered as the main source of the deviations observed in Fig. 3 from "ideal" behavior. Only a more complete analysis that fully takes account of the diffusion of precursors and adatoms (with the proper adatom GRAs, and involving the solution of differential diffusion equations) would be capable of fully describing the growth process in V-groove templates. This is carried out in the next section.

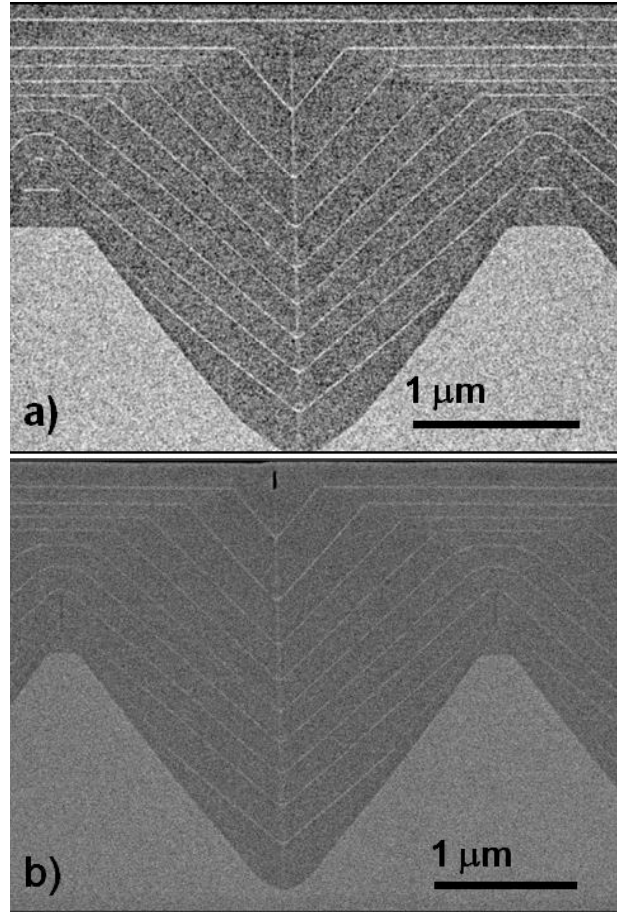

FIG. 4: Representative cross-sectional SEM images of the two samples analyzed in Fig. 3 (see text).

## IV. THEORY OF SELF-LIMITING GROWTH IN V-GROOVES

### A. Basic Formulation

Having highlighted the importance of decomposition rate anisotropies, we can now examine the atomistic mechanisms responsible for the formation of self-limited profiles. Our model is based on the following schematic MOVPE scenario. Trimethylgallium (TMGa), or trimethylaluminum (TMAl), and $AsH_3$ arrive at the substrate by diffusion through a boundary layer, after which these species and any fragments migrate with minimal lateral interactions. Decomposition reactions release Ga,

Al, and As preferentially at step edges[28], with a facet-dependent reaction rate. We will neglect the kinetics associated with the group-V species and assume that the facet growth rates are determined predominantly by the group-III species. This is justified in MBE growth of GaAs (001), which is carried out in the temperature range 550-650 °C under As-rich conditions, and typically As/Ga ≈ 10:1 and the Ga-As reaction kinetic are known not to be a limiting factor for growth[48]. Similarly in our MOVPE samples V/III ratios range from 100 to 800, and such high V/III ratios can only reinforce this assumption. The large feature sizes on the patterned substrate mean that direct simulations can be carried out only with the most advanced algorithms[49]. In such cases, a continuum formulation of the type pioneered by Burton, Cabrera, and Frank[50] provides a computationally less intensive alternative.

To simplify the description of our model, we consider a single migrating atomic species and we assume that Al and Ga adatoms are independent of one another, so that their effect in the growth of an alloy can be summed in proportion to their concentrations. The central quantity in our formulation of facet growth is the surface current $\mathbf{J}_i$ of adatoms at position $\mathbf{x}$ on the $i$th facet, which is given by Fick's first law:

$$\mathbf{J}_i = -D_i \boldsymbol{\nabla} n_i, \qquad (2)$$

where $D_i$ is the adatom diffusion constant and $n_i$ is the adatom concentration. The diffusion constant has the Arrhenius form[51]

$$D_i = a^2 \nu \exp\left(-\frac{E_i}{k_B T}\right), \qquad (3)$$

in which $a$ is the jump length, typically taken as the nearest neighbor lattice spacing, $\nu \sim 10^{13}\,\mathrm{s}^{-1}$ is the attempt frequency, and $E_i$ is the energy barrier to hopping. A geometrical prefactor can be included to account for the number of equivalent target sites of a mobile adatom.

The adatom concentration depends on three competing factors. (i) The deposition flux $F_i$ onto the $i$th facet. This process originates with the polyatomic precursors, which release the migrating atomic species through a decomposition reaction sequence. This step can be included as a separate set of equations[41] for the precursor kinetics. (ii) The adatom flux $\mathbf{J}_i$ across the substrate, which is driven by the concentration differences across the substrate. (iii) The lifetime $\tau_i$ of an adatom, which is the time to incorporation into the growth front from its release in the decomposition reaction. The equation for the evolution of the adatom concentration that includes these effects is obtained from Fick's second law as

$$\frac{\partial n_i}{\partial t} + \boldsymbol{\nabla} \cdot \mathbf{J}_i = F_i - \frac{n_i}{\tau_i}. \qquad (4)$$

As we are working at a single temperature here, the effect of the decomposition reaction can be replaced by an effective *adatom* flux on each facet and temperature-dependence of the adatom lifetime can be neglected. However, in the next section, we will discuss the effect of allowing the $\tau_i$ to vary with temperature.

We pause here for a few words concerning the relation between the approach we take here and previous theories of growth on patterned substrates. We are working at the level of individual atoms, so the quantities that distinguish between different facets are the effective adatom fluxes $F_i$ resulting from different precursor decomposition rates, the lifetimes to incorporation $\tau_i$, and the energy barriers $E_i$ between adjacent equilibrium positions on the surface. In particular, we need not include a separate chemical potential due to the surface energies of surrounding facets as the microscopic origins of such effects are subsumed in the $E_i$. There may be additional barriers to migration near the boundaries of two facets[52–54], but we have omitted these to avoid a proliferation of fitting parameters. Similarly we avoid introducing other non-kinetic terms such as entropy of mixing[21], which may play a role during equilibration, but are not expected to be important during growth.

To make the solutions of the coupled reaction-diffusion equations (4) analytically tractable, we will consider variations of the concentrations only along the $x$-direction, as shown in Fig. 5. Moreover, as we are concerned primarily with the self-limiting width in the V-groove, which is attained in the steady state, we will focus on the stationary solutions of these equations. Thus, when combined with (2), we have

$$D_i \frac{d^2 n_i}{dx^2} + F_i - \frac{n_i}{\tau_i} = 0 \qquad (5)$$

on each facet. This equation is supplemented by matching conditions at the facet boundaries to ensure the continuity of the adatom concentration and flux. The general solution to Eq. (5) is

$$n_i(x) = F_i \tau_i + A_i e^{-x/\lambda_i} + B_i e^{x/\lambda_i}, \qquad (6)$$

where $\lambda_i = (D_i \tau_i)^{1/2}$ and $A_i$ and $B_i$ are arbitrary constants. The quantity $F_i \tau_i$ represents the average concentration of the deposited adatoms during the time to incorporation, and $\lambda_i$ is the diffusion length of adatoms in the $i$th facet, i.e. the average distance an adatom travels prior to incorporation.

The constants $A_i$ and $B_i$ appearing in the solutions are determined by mandating the continuity of each concentration and the associated diffusive current across each facet boundary. The growth rate perpendicular to the $i$th facet at $x$ is obtained from these solutions as

$$\frac{dz_i}{dt} = \frac{\Omega_0}{\tau_i} n_i(x). \qquad (7)$$

The solution of this equation from the initial patterned substrate yields the evolution of the surface profile. The generation and manipulation of the solutions were carried out with MATHEMATICA.[55]

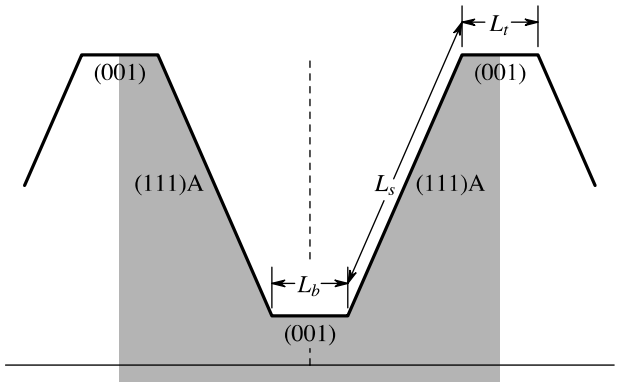

FIG. 5: One-dimensional V-groove profile showing three facets, the (001) at the top and bottom, and the (111)A along the sidewalls. The lengths of the top, side, and bottom facets are $L_t$, $L_s$, and $L_b$, respectively. The origin of the spatial coordinate $x$ is at the center of the bottom facet. The shaded region is the unit cell of the repeating pattern.

### B. Stationary Solutions for a V-Groove I

We first consider a patterned substrate with a periodic arrangement of (001) and (111)A facets, as shown in Fig. 5. At this stage we neglect the formation of any new facets during growth and consider only the configuration of alternating facets shown in the figure, but we will return to this question in the next section. For the simultaneous deposition of Ga and Al precursors according to the nominal concentration of the alloy, i.e. for $Al_xGa_{1-x}As$, the deposition flux consists of a fraction $x$ of TMAl precursors and a fraction $1-x$ of TMGa precursors. The precursors have different decomposition rates on the different facets of the substrate, which leads to different effective Al and Ga adatom fluxes on these facets. This is one of the factors leading to GRAs in our model.

Referring to Fig. 5, the effective deposition fluxes $F_b$, $F_s$, and $F_t$ of either adatom species on the bottom, side, and top facets are related to the overall growth rate of $F$ of the surface by considering deposition within the unit cell of the patterned substrate:

$$L_b F_b + 2L_s F_s + L_t F_t = (L_b + 2L_s + L_t)F. \quad (8)$$

We now stipulate that the top and bottom facets, each of which are (001) crystallographic faces, have the same effective fluxes, $F_t = F_b$, and introduce the ratio $r = F_s/F_b$ of the fluxes on side to the bottom facets. We thereby obtain

$$F_b = \frac{(L_b + 2L_s + L_t)F}{L_b + L_t + 2rL_s}, \quad (9)$$

with $F_s = rF_b$. For $Al_xGa_{1-x}As$, the Al and Ga fluxes are given by

$$F_b^{(Al)} = xF_b, \qquad F_b^{(Ga)} = (1-x)F_b, \quad (10)$$

$$F_s^{(Al)} = xr^{(Al)}F_b, \qquad F_s^{(Ga)} = (1-x)r^{(Ga)}F_b. \quad (11)$$

Upon fixing the overall deposition flux, which we take as 1 ML/s, $r^{(Al)}$ and $r^{(Ga)}$ completely specify the effective deposition fluxes onto the facets of the substrate.

The self-limiting profile will be calculated in terms of the *average* concentrations on each facet for each migrating species, rather than from the full spatially dependent profile (7) generated by solutions of the diffusion equations[56,57]. If we denote the average growth rate of the $i$th facet ($i = b, t, s$) due to the $k$th species ($k = $ Al, Ga) by $R_i^{(k)}$, then, from Eq. (7), we have that

$$R_i^{(k)} = \frac{1}{L_i}\int_i \frac{dz_i}{dt}\,dx = \frac{\Omega_0}{\tau_i^{(k)}}\bar{n}_i^{(k)}, \quad (12)$$

in which the integral is taken over the $i$th facet,

$$\bar{n}_i^{(k)} = \frac{1}{L_i}\int_i n_i^{(k)}(x)\,dx \quad (13)$$

is the average of the $k$th species on that facet, and $\tau_i^{(k)}$ the corresponding lifetime. The self-limiting width $L_b^*$ of the V-groove is obtained by the equating the *total* growth rates of the bottom and side facets:

$$\left[R_b^{(Al)} + R_b^{(Ga)}\right]\Big|_{L_b=L_b^*} = \left[R_s^{(Al)} + R_s^{(Ga)}\right]\Big|_{L_b=L_b^*}. \quad (14)$$

The solutions for the patterned surface in Fig. 5 are readily obtained. In fact, since $L_s \gg L_b$ and $L_s \gg L_t$, and we are concerned mainly with the growth kinetics near the bottom facet, we can neglect the top facet altogether. Moreover, the symmetry of the pattern means that we can restrict ourselves to the solution (6) along the bottom and one of the side facets. We obtain

$$n_b(x) = F_b\tau_b + A_b\cosh(x/\lambda_b), \quad (15)$$

for the bottom facet, for which $-\tfrac{1}{2}L_b \leq x \leq \tfrac{1}{2}L_b$. For the right side facet, we have

$$n_s(x) = F_s\tau_s + A_s e^{-x/\lambda_s}, \quad (16)$$

for which

$$\tfrac{1}{2}L_b \leq x \leq \tfrac{1}{2}L_b + L_s, \quad (17)$$

with the corresponding solution for the left side facet obtained by making the replacement $x \to -x$. The two constants $A_b$ and $A_s$ in these solutions are obtained from the continuity of the concentration and the adatom flux at the facet boundary,

$$n_b\!\left(\tfrac{1}{2}L_b\right) = n_s\!\left(\tfrac{1}{2}L_b\right), \quad (18)$$

$$D_b n_b'\!\left(\tfrac{1}{2}L_b\right) = D_s n_s'\!\left(\tfrac{1}{2}L_b\right). \quad (19)$$

Apart from the parametric dependence on various kinetic rates, the main dependence of $L_b^*$ is on the concentrations of Al and Ga, which are included in the growth rates through their effective fluxes. There are two types of parameters that enter our theory. First, there are quantities that are determined by the growth conditions, including the number and type of exposed facets on the substrate, their dimensions, and the substrate temperature (Table I). Each of these can be measured directly and thus cannot be varied to fit the experimentally determined limiting widths.

The other types of parameters describe the deposition and migration kinetics of the Al and Ga adatoms. Although the flow parameters of the precursors within the reactor are known, the kinetic parameters for the decomposition reaction(s) that release the group-III adatoms are not well characterized, so the quantities $r^{(k)}$ that characterize the relative deposition rates are treated as adjustable parameters. The kinetic parameters for the surface species are the diffusion barriers for the $k$th adatom on the $i$th facet, $E_i^{(k)}$, and the associated lifetimes to incorporation, $\tau_i^{(k)}$. For the bottom and side facets, we thus have four diffusion barriers and four lifetimes as adjustable parameters. There are altogether ten adjustable parameters but, as we discuss below, the experiments place stringent conditions on their values. The optimized parameters are compiled in Table II.

TABLE I: The experimental parameters used in Eq. (14) to fit the self-limiting widths in Fig. 6.

| Parameter | Value |
|---|---|
| $a$ | $2.71 \times 10^{-10}$ m |
| $L_s$ | $5500a = 1.49 \times 10^{-6}$ m |
| $L_t$ | $75a = 2.03 \times 10^{-8}$ m |
| $T$ | 973 K |

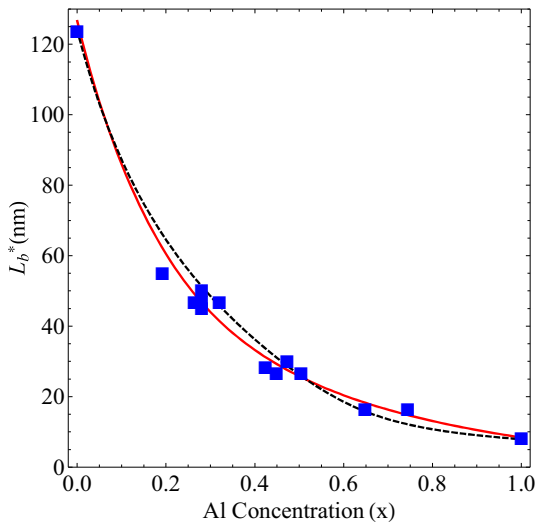

FIG. 6: (Color online) Comparison between the self-limiting width $L_b^*$ calculated from Eq. (14) with the parameters in Tables I and II (black, dashed line), the self-limiting width calculated with the model including (001) and (311)A facets (red, solid line - see Sec. IV C ) and the experimental data (blue squares) in Ref. [21].

TABLE II: Kinetic parameters for Ga and Al adatoms used in Eq. (14) to fit the self-limiting widths in Fig. 6 (black, dashed line).

| Parameter | Al | Ga |
|---|---|---|
| $E_b^{(k)}$ | 2.15 eV | 1.90 eV |
| $E_s^{(k)}$ | 1.40 eV | 1.0 eV |
| $\tau_b^{(k)}$ | 0.70 s | 1.25 s |
| $\tau_s^{(k)}$ | 1.0 s | 1.75 s |
| $r^{(k)}$ | 1.6 | 1.1 |

The results of our calculations are compared with measured self-limiting widths in Fig. 6. The trend produced by our theory is in excellent agreement with the experimental data points reported in Ref. [21]. The diffusional energy barriers indicate that the Al adatoms are appreciably less mobile than Ga atoms, which conforms to general expectations based on *ab initio* density functional calculations[58]. However, these barriers are strong functions of the surface reconstruction on the facet[59], which is not readily accessible in these experiments, so we must be cautious about interpreting our values as quantitative estimates of atomistic rates. Although the agreement between our theory and experiment in Fig. 6 is encouraging, an important inaccuracy of the model is its inability to predict the correct Ga concentration profiles. As Fig. 7 shows, there is Ga segregation near the center of the (100) bottom facet, which the model does not fully reproduce, despite accounting for the behavior of the concentration-dependence of $L_b^*$. This indicates that the self-limiting width is not in itself a reliable indication of the concentration profile in the bottom of the V-groove and that the spatial distribution of the growth rates provides a more critical assessment of our model. In the next section, we will consider refinements to our basic model that provide significant improvement to the concentration profiles.

### C. Stationary Solutions for a V-Groove II

The basic approach described in the previous section provides a framework for the improvement of our model. Figure 7 shows a schematic illustration of the arrangement of facets seen in experiments[22], where the bottom (001) facet is bounded by (311)A facets, and the associated segregation of Ga to the (001) and (311)A facets[22]. Including the (311)A facet within our computational framework is straightforward, with the only substantial difference arising from the fact that the self-limiting width must include both the base (001) and (311)A facets:

$$L_b^* = L_b + L_{(311)A}. \tag{20}$$

There are now additional parameters for the kinetics on the (311)A facet, but there are guidelines for choosing the kinetic parameters for this facet. Due to the crystal structure, (311)A is less corrugated than (100) but more than the (111)A side facet, so the energy barriers are expected to be lower than on the bottom but larger than on the side. The opposite applies to the incorporation times. There are now two growth rate ratios

$$F_s^{(k)} = r_s^{(k)} F_b^{(k)}, \qquad F_{(311)A}^{(k)} = r_{(311)A}^{(k)} F_b^{(k)}, \qquad (21)$$

with $r_s^{(k)} > r_{(311)A}^{(k)}$, so the equations of Sec. IV B have to be adjusted accordingly. Finally, the condition that determines the self-limiting width is

$$\left[ R_b^{(Al)} + R_b^{(Ga)} \right]\Big|_{L_b=L_b^*} = \left[ R_{(311)A}^{(Al)} + R_{(311)A}^{(Ga)} \right]\Big|_{L_b=L_b^*} \qquad (22)$$

$$= \left[ R_s^{(Al)} + R_s^{(Ga)} \right]\Big|_{L_b=L_b^*}. \qquad (23)$$

To solve these equations, we used the kinetic parameters compiled in Tables I and III. The results of our analysis are shown in Figs. 6 and 8. The solid red curve in Fig. 6 shows the self-limited profile base as Al concentration changes in comparison with the data in Ref. [21] and the same profile calculated for the model including only the (001) bottom. Similarly to previous results there is a good agreement between experimental data and our calculation. As for the concentration profile, shown in Fig 8, the peaks in the center of the bottom (100) facet are present, in agreement with experiments, but the peak in the center on the (311)A facets is not accounted for. A possible cause of this behavior is an additional barrier to migration between the (311)A and other facets. Theoretical studies[60] on GaAs surfaces based on a combination of empirical interatomic potentials and *ab initio* calculations have revealed that Ga preferentially adsorbs on the (311)A facet and there is a preferential migration from the (001)-$\beta 2(2\times 4)$ to the (311)A facets. The incorporation of such effects is a matter for future work.

TABLE III: Kinetic parameters for Ga and Al adatoms used in Eq. (22) and (23), for the model including the (311)A oriented facets, to fit the self-limiting widths in Fig. 6 (red, solid line).

| Parameter | Al | Ga |
|---|---|---|
| $E_b$ | 2.15 eV | 1.90 eV |
| $E_{(311)A}$ | 1.55 eV | 1.55 eV |
| $E_s$ | 1.45 eV | 1.30 eV |
| $\tau_b$ | 0.70 s | 1.35 s |
| $\tau_{(311)A}$ | 0.99 s | 1.44 s |
| $\tau_s$ | 1.0 s | 1.50 s |
| $r_s$ | 1.6 | 1.1 |
| $r_{(311)A}$ | 1.01 | 1.01 |

## V. INTERPLAY BETWEEN PRECURSOR DECOMPOSITION ANISOTROPIES AND CAPILLARITY FLUXES IN V-GROOVES

In the preceding section we focussed on the formation of a self-limiting profile near the bottom of V-grooves at a single temperature. This simplified the optimization of parameters because the temperature-dependence of the decomposition and adatom lifetime could be neglected. In this section, we address this issue by using a coarse-grained reaction-diffusion equation, together with experimental observations, to explain the origin of the temperature-dependence of angle between the (111)A sidewalls with respect to the basal facet.

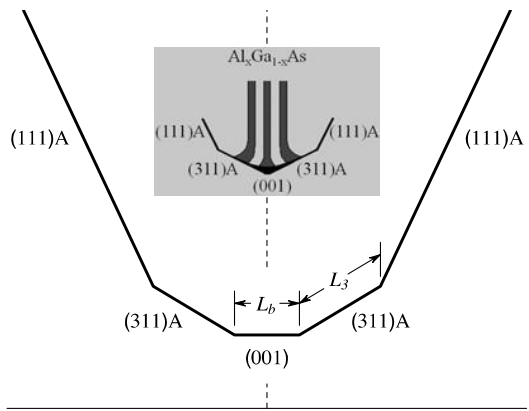

FIG. 7: Schematic illustration of the facets near the bottom of a V-groove, as seen in experiments, with the bottom (001) facets bounded by (311)A and (111)A facets. The length of the (311)A facet is denoted by $L_3$. Inset: The self-limiting concentration profiles on the (001) and (311)A oriented facets on the bottom. Segregation in AlGaAs leads to the formation of three narrow Ga enriched branches in the middle of each plane.

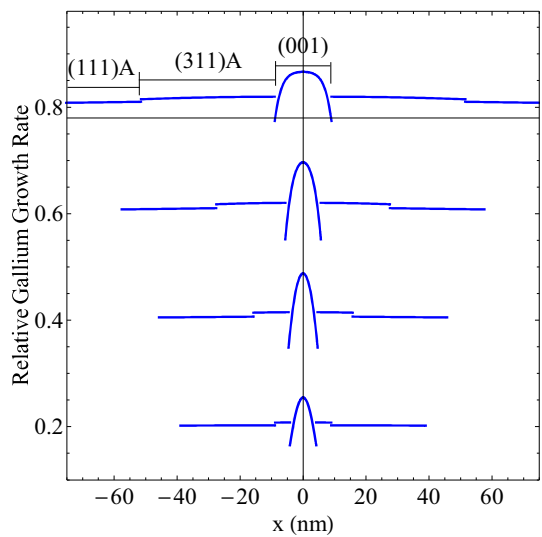

FIG. 8: Calculated steady-state Ga concentration for a model that includes growth on the (011), (311)A and (111)A facets, which are labeled for the entry for a Ga concentration of 0.8.

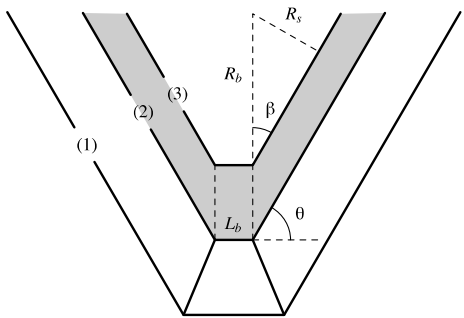

FIG. 9: Schematic cross-section of the bottom of a V-groove sample after the growth of $Al_xGa_{1-x}As$. Three layers are sketched. From an initial broad bottom profile (1) the system reaches a self-limited profile (2). Once this profile is reached, for the subsequent layers (3), the bottom width $L_b$ and the angles $\theta$ and $\beta$ remain unchanged for a given $Al_xGa_{1-x}As$ concentration. The ratio between the growth rates $R_b$ [bottom (100) facet] and $R_s$ [vicinal (111)A facet] is fixed by the geometric constraints.

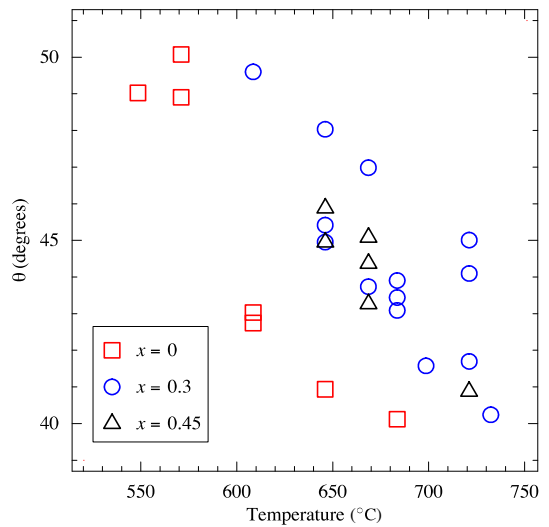

FIG. 10: Variation of the vicinal (111)A sidewall planes angle $\theta$ as a function of growth temperature in $Al_xGa_{1-x}As$, with $x = 0$ (squares), $x = 0.3$ (circles) and $x = 0.45$ (triangles). Temperature values have been renormalized from those reported in Ref. [22] to show the estimated real temperature.

The angle $\theta$ formed between the vicinal (111)A plane and the (100) growth surface (Fig. 9) varies from $\sim 40°$ to $\sim 50°$ when the growth temperature is varied from $\sim 500°$C [resp., $\sim 600°$C] to $\sim 650°$C [resp., $\sim 750°$C] for GaAs [resp., $Al_{0.30}Ga_{0.70}As$][22]. For convenience, the data from Ref. [22] (see Fig. 9 for definitions) are shown in Fig. 10.[61]

We caution the reader that these data are quite scattered when the AlGaAs alloys are considered. These are the only data available in the literature with internal consistency (constant V-groove and growth parameters, only modified sample temperature), and we report them here for clarity as they show anyway a clear trend. This is the only thing we discuss qualitatively, refraining in the following from a quantitative analysis.

We first observe that once a self-limited profile is reached (layer 3 in Fig. 9)[62] the growth rate $R_b$ of the bottom (100) facet and the growth rate $R_s$ of the vicinal (111)A facet (both chosen here, as before, vertical to their respective growth plane) cannot be independent, but are linked by a simple geometrical relation: $R_s/R_b = \sin\beta = \cos\theta$. This has the important consequence that any physical mechanism producing a "flattening" of the plane angle with increasing growth temperatures must increase the growth rate of the (111)A vicinal facets relative to that of the bottom (100) facet. At first sight this might seem to contradict the discussion in Sec. II, where we indicated a higher growth rate on the (100) ridge facet at higher temperatures as a result of the competition in the decomposition process. Nevertheless, $R_b$ cannot be treated in the same way as the growth rate on the (100) ridge planes.

There are, in fact, several physical processes that contribute to the overall growth rate which do not have an important role for the ridge facets, but cannot be altogether neglected:[21]

A. The precursor decomposition process clearly contributes to $R_b$, but the bottom (100) facet is relatively limited in dimensions (20-100 nm) and the amount of material deposited is largely overcome by a second, and more relevant physical process,

B. Capillarity,[20–22] which induces a strong adatom flux from the side facets to the bottom facet, as discussed in Ref. [21].[63] Other physical processes cannot be altogether neglected:

C. The anisotropy in the decomposition process and the capillarity fluxes will induce an anisotropy in the free adatom concentrations on the vicinal (111)A facets relative to the (100) facet. This concentration difference between the two planes will necessarily imply an adatom flux that will tend to reduce the differences between the free adatom concentration on the (100) and the (111)A vicinal facets.

D. In the case of $Al_xGa_{1-x}As$ alloys, a fourth term (D) has been proposed: an entropy of mixing contribution.[21] We will neglect here this term as it should not be expected to have a major role on $\theta$. This term could be present in fact only in the $Al_xGa_{1-x}As$ case, and the overall physical process appears to be similar for both GaAs (no entropic effects possible) and AlGaAs.

We will show below that, if we limit the discussion to the most relevant flux term determining $R_b$, i.e., the capillarity induced term (B) (which is, in the temperature range of interest, 2-3 times higher than the others, as derived in Ref. [21]), we expect to observe, at least qualitatively, the trends in Fig. 10. We will, for completeness, treat the discussion of the other terms at the end of this section.

Consider a spatially coarse-grained description of the time-dependent concentration $n_b$ on the bottom facet. We include three terms: (i) the effective deposition flux $F_b$, (ii) the depletion due to incorporation, which yields the term $-n_b/\tau_b$, and (iii) source terms from the two (111) side facets, which we determine from the diffusion current density in (2):

$$J = D_s \frac{dn_s}{dx} \approx \frac{D_s \bar{n}_s}{\lambda_s}. \tag{24}$$

Here, $\bar{n}_s$ is the average adatom concentration on the side-facet, $D_s$ the corresponding diffusion constant, and $\lambda_s$ the diffusion length. Hence, the evolution equation for $n_b$ is

$$\frac{dn_b}{dt} = F_b + \frac{D_s \bar{n}_s}{\lambda_s}\left[\delta\left(x+\tfrac{1}{2}L_b\right) + \delta\left(x-\tfrac{1}{2}L_b\right)\right] - \frac{n_b}{\tau_b}, \tag{25}$$

where the $\delta$-functions indicate source terms from the two edges of the bottom facet. By integrating from $-\tfrac{1}{2}L_b - \epsilon$ to $\tfrac{1}{2}L_b + \epsilon$ ($\epsilon$ is a small quantity that ensures that the integration captures the $\delta$-functions and is set to zero after the calculation) and dividing the result by $L_b$, we obtain the following equation for the average concentration $\bar{n}_b$ of the bottom facet:

$$\frac{d\bar{n}_b}{dt} = F_b + \frac{2 D_s \bar{n}_s}{\lambda_s L_b} - \frac{\bar{n}_b}{\tau_b}. \tag{26}$$

Using the fact that $\lambda_s = (D_s \tau_s)^{1/2}$, the second term on the right-hand side of this equation can be written as $\lambda_s \bar{n}_s / L_b \tau_s$. Then, in the steady state, we have that

$$\frac{\bar{n}_b}{\tau_b} = \frac{2\lambda_s \bar{n}_s}{L_b \tau_s} + F_b. \tag{27}$$

If we consider a temperature regime where the dominant source of material to the bottom facet is from the side facets, rather than the effective flux,

$$\frac{\bar{n}_b}{\tau_b} \approx \frac{2\lambda_s \bar{n}_s}{L_b \tau_s}, \tag{28}$$

which, by invoking (12), immediately yields

$$R_b \approx \frac{2\lambda_s}{L_b} R_s, \tag{29}$$

or,

$$\frac{R_b}{R_s} \approx \frac{2\lambda_s}{L_b} = \frac{2(D_s \tau_s)^{1/2}}{L_b}. \tag{30}$$

To proceed further, we must identify the temperature dependence of each quantity on the right-hand side. The diffusion constant has already been taken to have the Arrhenius form in (3):

$$D_s = a^2 \nu \exp\left(-\frac{E_s}{k_B T}\right). \tag{31}$$

The incorporation lifetime can be assigned a similar temperature dependence:

$$\tau_s = \nu^{-1} \exp\left(\frac{E_\tau}{k_B T}\right), \tag{32}$$

where, in the interest of simplicity, we have used the same frequency prefactor as for the diffusion constant. Finally, $L_b$ has been reported to vary approximately as[21,22]

$$L_b \simeq L^* \exp\left(-\frac{E_b}{3 k_B T}\right). \tag{33}$$

Hence,

$$\frac{R_b}{R_s} \approx \frac{a}{L^*} \exp\left(\frac{E_\tau - \tfrac{1}{6} E_s}{k_B T}\right). \tag{34}$$

If $E_\tau - \tfrac{1}{6} E_s > 0$, then $R_b/R_s$ is a decreasing function of the temperature, which implies the $\cos\theta$ increases with $T$, which explains the trend in Fig. 10.

We now consider the value of $E_\tau$. The incorporation process subsumes adatom motion and the morphology of a facet. In the simplest case, where an adatom arrives a facet, diffuses to a step edge and incorporates there, the energy barrier is the same at that of diffusion. If, however, the adatom undergoes several cycles of attachment and detachment prior to incorporation, then the effective energy barrier *increases* because the detachment barriers from step edges are *greater* than those for adatom hopping on a terrace. Hence, the requirement that $E_\tau - \tfrac{1}{6} E_s > 0$ is expected to be fulfilled.

We also note that the decreasing exponential dependence of $\theta$ on $T$ also correctly predicts the tendency of the $\theta$ to saturate at high temperatures, at least if $E_\tau - \tfrac{1}{6} E_s > 0.5$ eV. Nevertheless, this functional dependence cannot correctly predict a similar saturation at low temperatures, where the angle gradients with $T$ seem also to decrease. We observe that this discrepancy appears in a temperature range (e.g. $T_g < 550°$C for GaAs, $<600$ °C for $Al_{0.3}Ga_{0.7}As$), where the capillarity term is known to become small or negligible,[64] and that the system is likely to enter a regime of kinetically limited processes as opposed to the diffusion limited processes at higher $T$.[36] Hence, it is not surprising that our simplified assumptions on the epitaxial process start to be inaccurate at these temperatures.

For completeness we will now briefly review the other terms contributing to $R_b$. We first observe that if we allow ourselves to treat them discretely, then, $R_b \simeq A + B + C + D$ and we will have that

$$\frac{1}{\cos\theta} = \frac{R_b}{R_s} \simeq \frac{A}{R_s} + \frac{B}{R_s} + \frac{C}{R_s} + \frac{D}{R_s}. \tag{35}$$

As already mentioned, the dependence of $\theta$ on the fourth term should not be significant. The third flux term has a functional dependence of the form $(D_0/R_s)\exp(-E_b/k_B T)\partial n/\partial x$ and will depend significantly on $\partial n/\partial x$. This term will in general tend to re-equilibrate the adatom concentration of the (111)A and (100) facets, by partially counteracting the capillarity term when a strong capillarity flux is present. There is no reason to assume this should have a major role on the relative evolution of $R_b$ and $R_s$, but should simply contribute to damping the variations induced by the capillarity term without modifying trends.

The term $A/R_s$ has the form

$$\frac{R_{(100)}}{R_{(111)}} \simeq A\exp\left[\frac{E_{(111)} - E_{(100)}}{k_B T}\right], \qquad (36)$$

as discussed in the previous paragraph. This opposes the flattening of $\theta$ with $T$. As already discussed, the reduced (100) bottom facet dimensions and all experimental evidence (the formation of quantum wires in primis, but also the presence of a segregated vertical quantum well) suggest that $R_b$ contributed mainly to the capillarity term, which is expected to be a few times larger than the others.[21] The only speculation possible here is that the gradients in the first term are not sufficient to compensate the capillarity induced trend, especially in the range of temperatures where the maximum variations of $\theta$ are observed. Again only a proper determination (at the moment impractical) of all the terms involved (both prefactors and potential barriers) and the exact solution of the differential diffusion equations will clarify if our hypotheses are correct.

## VI. CONCLUSIONS

We have clarified that anisotropies in the precursor decomposition process on the different crystallographic planes are the main reason for the observed growth rate anisotropies and self-limited profiles during seeded MOVPE of nanostructures on pre-patterned semiconductor substrates. We have demonstrated that the evolution of the nanostructures self-limited profile relies explicitly on kinetic terms. Moreover, we showed that it is possible to qualitatively describe the evolution of (111)A vicinal facet formation with growth temperature in V-grooved substrates as a result of the combined evolution with temperature of the anisotropic decomposition processes and of the capillarity fluxes towards the V-groove bottom (100) facet. Future work will focus on modelling the effect of temperature on the anisotropic decomposition rates and adatom diffusion, with the goal to better support and corroborate the qualitative study here presented. We believe our findings will have a substantial impact on the knowledge and control over the epitaxial process of seeded nanostructures, improving structural quality and enhancing transport and optical properties of V-groove QWRs and pyramidal QDs.

## VII. ACKNOWLEDGEMENTS

We would like to thank S. Mautino for the AFM images in Fig. 2 and G. Biasiol for the data of Fig. 10. We are also grateful to G. Schusteritsch, S. Watanabe, B. Dwir and A. Zangwill for useful discussions. This research was in part funded by Science Foundation Ireland under grant 05/IN.1/I25.

[1] G. S. Kar, S. Kiravittaya, M. Stoffel, and O. G. Schmidt, Phys. Rev. Lett. **93**, 246103 (2004).
[2] G. Springholz, J. Stangl, M. Pinczolits, V. Holy, P. Mikulik, P. Mayer, K. Wiesauer, G. Bauer, D. Smilgies, H. H. Kang, L. Salamanca-Riba, Physica E **7**, 870 (2000).
[3] A. Badolato, K. Hennessy, M. Atatüre, J. Dreiser, E. Hu, P. M. Petroff, A. Imamoğlu, Science **308**, 1158 (2005).
[4] K. Hennessy, A. Badolato, M. Winger, D. Gerace, M. Atatüe, S. Gulde, S. Fält, E. L. Hu, and A. Imamoğlu, Nature **445**, 896 (2007).
[5] B. D. Gerardot, S. Strauf, M. J. A. de Dood, A. M. Bychkov, A. Badolato, K. Hennessy, E. L. Hu, D. Bouwmeester, and P. M. Petroff, Phys. Rev. Lett. **95**, 137403 (2005).
[6] A. Hartmann, Y. Ducommun, L. Loubies, K. Leifer and E. Kapon, Appl. Phys. Lett. **73**, 2322 (1998).
[7] D. Kaufman, Y. Berk, B. Dwir, A. Rudra, A. Palevski, and E. Kapon, Phys. Rev. B **59**, R10433 (1999).
[8] L. Sirigu, D. Y. Oberli, L. Degiorgi, A. Rudra, and E. Kapon, Phys. Rev. B **61**, R10575 (2000).
[9] E. Levy, A. Tsukernik, M. Karpovski, A. Palevski, B. Dwir, E. Pelucchi, A. Rudra, E. Kapon, and Y. Oreg, Phys. Rev. Lett. **97**, 196802 (2006).
[10] E. Pelucchi, M. Baier, Y. Ducommun, S. Watanabe, and E. Kapon, Phys. Stat. Solidi B **238**, 233 (2003).
[11] M. H. Baier, S. Watanabe, E. Pelucchi, and E. Kapon, Appl. Phys. Lett. **84**, 1943 (2004).
[12] K. Leifer, E. Pelucchi, S. Watanabe, F. Michelini, B. Dwir and E. Kapon, Appl. Phys. Lett. **91**, 081106 (2007).
[13] S. Watanabe, E. Pelucchi, B. Dwir, M. Baier, K. Leifer, E. Kapon, Physica E **21**, 193 (2004).
[14] S. Watanabe, E. Pelucchi, B. Dwir, M. Baier, K. Leifer, and E. Kapon, Appl. Phys. Lett. **84**, 2907 (2004).
[15] L. O. Mereni, V. Dimastrodonato, R. J. Young, and E. Pelucchi, Appl. Phys. Lett. **94**, 223121 (2009).
[16] A. Mohan, P. Gallo, M. Felici, B. Dwir, A. Rudra, J. Faist, and E. Kapon, Small **6**, 1268 (2010).
[17] M. H. Baier, E. Pelucchi, E. Kapon, S. Varoutsis, M. Gallart, I. Robert-Philip, and I. Abram, Appl. Phys. Lett. **84**, 648 (2004).
[18] A. Mohan, M. Felici, P. Gallo, B. Dwir, A. Rudra, J. Faist, and E. Kapon, Nature Photonics **4**, 302 (2010).
[19] M. Baier, C. Constantin, E. Pelucchi, and E. Kapon, Appl. Phys. Lett. **84**, 1967 (2004).
[20] M. Ozdemir and A. Zangwill, J. Vac. Sci. Technol. A **10**, 684 (1992).
[21] G. Biasiol and E. Kapon, Phys. Rev. Lett. **81**, 2962 (1998); G. Biasiol, A. Gustafsson, K. Leifer, E. Kapon, Phys. Rev.

B **65**, 205306 (2002).
22. G. Biasiol, Ph.D. Thesis no 1859, *Formation Mechanisms of Low-Dimensional Semiconductor Nanostructures Grown by OMCVD on Nonplanar Substrates* (Swiss Federal Institute of Technology, Lausanne, 1998).
23. T. Sugaya, K.-Y. Jang, C.-K. Hahn, M. Ogura, K. Komori, A. Shinoda, and K. Yonei, J. Appl. Phys. **97**, 034507 (2005).
24. A. Hartmann, L. Loubies, F. Reinhardt, and E. Kapon, Appl. Phys. Lett. **71**, 1314 (1997).
25. T. Tsujikawa, W. Pan, K. Momma, M. Kudo, K. Tanaka, H. Yaguchi, K. Onabe, Y. Shiraki and R. Ito, Jpn. J. Appl. Phys. **36**, 4102 (1997).
26. S. H. Jones, L. K. Seidel, K. M. Lau, and M. Harold, J. Cryst. Growth **108**, 73 (1991).
27. F. Grosse and R. Zimmermann, J. Crystal Growth **212**, 128 (2000).
28. F. Lelarge, G. Biasiol, A. Rudra, A. Condo and E. Kapon, Microelectron. J. **30**, 461 (1999).
29. Similar comments can be found also in earlier literature, e.g. S. D. Hersee, E. Barbier, and R. Blondeau, J. Crystal Growth **77**, 310 (1986).
30. A. Kaluza, A. Schwarz, D. Gauer, H. Hardtdegen, N. Nastase, H. Luth, T. Schapers, D. Meertens, A. Maciel, J. Ryan, E. O'Sullivan, J. Crystal Growth **221**, 91 (2000).
31. E. Pelucchi, S. Watanabe, K. Leifer, B. Dwir, Q. Zhu, P. De Los Rios and E. Kapon, NanoLetters **7**, 1282 (2007).
32. E. Pelucchi, S. Watanabe, K. Leifer, B. Dwir and E. Kapon, Physica E **23**, 476 (2004).
33. S. Watanabe, E. Pelucchi, K. Leifer, A. Malko, B. Dwir, and E. Kapon, Appl. Phys. Lett. **86**, 243105 (2005).
34. See for example, the recent review by X.-L. Wang and V. Voliotis, J. Appl. Phys. **99**, 121301 (2006).
35. We stress that this is not a unique "sample." We observed this phenomenon on all the samples where areas with patterned pyramids are alternated with unpatterned regions.
36. The entire decomposition process is known to occur in three different stages. It is not our objective to discuss this point in detail. For a general overview of this issue see, for example, G. B. Stringfellow, *Organometallic Vapor-Phase Epitaxy: Theory and Practice* 2nd ed. (Academic Press, London, 1998).
37. Quantum cascade lasers have been grown by MBE on (111)B substrates. See, for example, J.-Y. Bengloan, A. De Rossi, V. Ortiz, X. Marcadet, M. Calligaro, I. Maurin and C. Sirtori, Appl. Phys. Lett. **84**, 2019 (2004).
38. T. Fukui and H. Saito, J. Crystal Growth **107**, 231 (1991).
39. T. Nishida and N. Kobayashi, Jpn. J. Appl. Phys. **37**, L13 (1998).
40. T. Shitara, T. Kaneko, and D. D. Vvedensky, Appl. Phys. Lett. **63**, 3321 (1993).
41. A. L.-S. Chua, E. Pelucchi, A. Rudra, B. Dwir, E. Kapon, A. Zangwill, and D. D. Vvedensky, Appl. Phys. Lett. **92**, 013117 (2008).
42. See for example I. Štich, A. De Vita, M. C. Payne, M. J. Gillan, and L. J. Clarke, Phys. Rev. B **49**, 8076 (1994).
43. D. Wüllner, M. Chahoud, T. Schrimpf, P. Bönsch, H.-H. Wehmann, and A. Schlachetzki, J. Appl. Phys. **85**, 249 (1999).
44. Gas phase effects would be important if we used atmospheric pressure conditions, which would result in a closing (and actually no real self limited profile of lateral facets) of the V-grooves, as gas phase diffusion would increase the growth rate at the top of the inverted V-groove (or pyramid) and reduce it at the bottom. Only for growth processes at atmospheric pressure our discussion should add these effects to the physics described in our manuscript.
45. Gas phase effects are obviously not expected to show such a drastic dependence on $T_g$.
46. For the sake of simplicity, we are again neglecting here the fact that TMGa/Al decompose through a multistep process. See Ref. [36].
47. We neglect adatom diffusion effects, and the adatom diffusion induced by differences in surface chemical potentials, which in this picture are only effective in "flattening" the crystallographic plane profile.
48. P. Chen, J.Y. Kim, A. Madhukar and N.M. Cho, J. Vac. Sci. Technol. B **4**, 890 (1986).
49. N. Haider, S. A. Khaddaj, M. R. Wilby and D. D. Vvedensky, Computers in Physics **9** 85 (1995).
50. W. K. Burton, N. Cabrera, and F. C. Frank, Phil. Trans. Roy. Soc. London **243**, 299 (1951).
51. A. Zangwill, *Physics at Surfaces* (Cambridge University Press, Cambridge, UK, 1988).
52. M. Hata, T. Isu A. Watanabe, and Y. Katayama, J. Vat. Sci. Technol. B **8**, 692 (1990).
53. M. Hata, A. Watanabe, and T. Isu, J. Cryst. Growth **111**, 83 (1991).
54. N. Haider, M. R. Wilby, and D. D. Vvedensky, Appl. Phys. Lett. **62**, 3108 (1993).
55. Wolfram Research, Inc., Urbana-Champaign, IL; v. 7.0.1.0.
56. T. Sato, I. Tamai, and H. Hasegawa, J. Vac. Sci. Technol. B **22**, 2266 (2004).
57. T. Sato, I. Tamai, and H. Hasegawa, J. Vac. Sci. Technol. B **23**, 1706 (2005).
58. A. Kley, P. Ruggerone, and M. Scheffler, Phys. Rev. Lett. **79**, 5278 (1997).
59. J. G. LePage, M. Alouani, D. L. Dorsey, J. W. Wilkins, and P. E. Blöchl, Phys. Rev. B **58**, 1499 (1998).
60. K. Asano, Y. Kangawab, H. Ishizakia, T. Akiyamaa, K. Nakamuraa, and T. Ito, Appl. Surf. Sci. **237**, 206 (2004).
61. A consistent dependence of $\theta$ with growth temperature is also observed in the samples analyzed in this paper.
62. For simplicity, we have neglected the actual faceting effects at the bottom of the V-groove and, in particular, the (311) forming at each side of the bottom (100) facet.
63. The effect of capillarity is, due to geometric reasons, even stronger in the case of pyramidal structures. See, for example, Q. Zhu, E. Pelucchi, S. Dalessi, K. Leifer, M.-A. Dupertuis and E. Kapon, Nano Lett. **6**, 1036 (2006).
64. F. Lelarge, C. Constantin, K. Leifer, A. Condo, V. Iakovlev, E. Martinet, A. Rudra, and E. Kapon, Appl. Phys. Lett. **75**, 3300 (1999).